\documentclass[aps,twocolumn]{revtex4}
\begin{document}
\title{On the vector solutions of Maxwell equations \\ with the
spin-weighted spherical harmonics}
\author{E.A. Matute}
\affiliation{Departamento de F\'{\i}sica, Universidad de Santiago de
Chile \\ Casilla 307 - Correo 2, Santiago, Chile \\
e-mail: ematute@lauca.usach.cl}
\begin{abstract}
\noindent The discussion of our recent work concerning the vector
solution of boundary-value problems in electromagnetism is
extended to the case of no azimuthal symmetry by means of the
spin-weighted spherical harmonics.
\\[7pt]
\\[10pt]
\noindent Se extiende la discusi\'on de nuestro trabajo reciente
sobre la soluci\'on vectorial de problemas con valores de frontera
en electromagnetismo al caso sin simetr\'{\i}a azimutal mediante
el uso de los arm\'onicos esf\'ericos con peso de esp\'{\i}n.
\\[7pt]
\emph{Descriptores:} Ecuaciones de Maxwell; campos el\'ectrico y
magn\'etico; problemas con valores de frontera; arm\'onicos
esf\'ericos con peso de esp\'{\i}n.
\\[10pt]
\end{abstract}
\maketitle

In a recent paper~\cite{Matute}, we introduced a somewhat
different approach to solving boundary-value problems in spherical
coordinates for time-independent and time-dependent electric and
magnetic fields, without involving the scalar and vector
potentials. We showed that the process includes the same
mathematics of separation of variables as the usual approach of
solving for potentials. However, it is restricted by the
assumption of azimuthal symmetry.  We now wish to remove this
constraint. So the natural complete orthonormal set of expansion
functions to consider for the vector solutions of Maxwell
equations are the spin-weighted spherical
harmonics~\cite{Scanio,Brooks,Torres1,Torres2}.  The purpose of
this note is to recast the general vector solutions for the
electric and magnetic fields and illustrate their applications on
boundary-value problems by dealing, for simplicity's sake, with
the commonplace examples solved in Ref.~\cite{Matute}.

For time-independent electric and magnetic phenomena, the fields
outside sources satisfy the vector Laplace equation $\nabla^{2}
{\bf F} = {\bf 0}$ with the subsidiary condition ${\bf \nabla}
\cdot {\bf F} = 0$.  The general solution in terms of the
spin-weighted spherical harmonics $_{s}Y_{lm}$ \cite{Torres1},
with spin weight $s=0, \pm 1$ and $_{0}Y_{lm}=Y_{lm}$, can be
written as
\begin{eqnarray}
F_{0} & = & \sum_{l=0}^{\infty} \sum_{m=-l}^{l}  \left( a_{lm}
r^{l-1} + \frac{b_{lm}}{r^{l+2}} \right) Y_{lm} , \nonumber \\
F_{\pm} & = & \sum_{l=1}^{\infty} \sum_{m=-l}^{l} v_{\pm} \;\;
_{\pm 1}Y_{lm} ,
\label{Laplace}
\end{eqnarray}
with
\begin{eqnarray}
v_{\pm} & = & c_{lm} r^{l} + \frac{d_{lm}}{r^{l+1}} \nonumber
\\ & & \mp \sqrt{\frac{l(l+1)}{2}} \left( \frac{a_{lm}}{l} \; r^{l-1} -
\frac{b_{lm}}{l+1} \; \frac{1}{r^{l+2}} \right) ,
\label{v}
\end{eqnarray}
where $F_{0}={\bf e}_{0}\cdot{\bf F}=F_{r}$, $F_{\pm}={\bf
e}_{\pm} \cdot{\bf F}=(F_\theta \pm i F_{\varphi}) / \sqrt{2}$ are
the components of the field with spin weight $s=0,\pm 1$,
respectively, and ${\bf e}_{0}=\hat{\bf r}$, ${\bf
e}_{\pm}=(\hat{\bf \theta} \pm i \hat{\bf \varphi}) / \sqrt{2}$
are the spin-weighted combinations of the orthonormal basis.

For harmonic time-dependent sources and fields, the electric and
magnetic fields in regions apart from sources satisfy the vector
Helmholtz equation $\nabla^{2}{\bf F}+k^2{\bf F}={\bf 0}$ with the
transverse condition ${\bf \nabla}\cdot{\bf F}=0$. The general
solution \cite{Scanio,Torres1} now becomes
\begin{eqnarray}
F_{0} & = & \sum_{l=0}^{\infty} \sum_{m=-l}^{l}  \left[ a_{lm}
\frac{j_{l}(kr)}{r} +
b_{lm} \frac{n_{l}(kr)}{r} \right] Y_{lm} , \nonumber \\
F_{\pm} & = & \sum_{l=1}^{\infty} \sum_{m=-l}^{l} w_{\pm} \;\;
_{\pm 1}Y_{lm} ,
\label{Helmholtz}
\end{eqnarray}
with
\begin{eqnarray}
w_{\pm} & = & c_{lm} j_{l}(kr) + d_{lm} n_{l}(kr) \nonumber \\
& & \displaystyle \mp \frac{a_{lm}}{\sqrt{2l(l+1)}} \frac{1}{r} \;
\frac{d\;}{dr} [r \, j_{l}(kr)] \nonumber \\ & & \displaystyle \mp
\frac{b_{lm}}{\sqrt{2l(l+1)}} \frac{1}{r} \; \frac{d\;}{dr} [r \,
n_{l}(kr)] ,
\label{w}
\end{eqnarray}
where the spherical Hankel functions, $h^{(1)}_{l}$ and
$h^{(2)}_{l}$, instead of the spherical Bessel functions, $j_{l}$
and $n_{l}$, may be required by boundary conditions.

We remark that $F_{+}=F_{-}$ in the case of boundary-value
problems having azimuthal symmetry with $F_{\varphi}=0$. This
implies that $c_{lm}=d_{lm}=0$ in Eqs. (\ref{v}) and (\ref{w}),
which leads in turn to the solutions obtained in
Ref.~\cite{Matute}.

The boundary conditions for the electric and magnetic fields must
be expressed in terms of their spin-weighted components. Assuming
that the boundary surface is a sphere with ${\bf n}={\bf e}_{0}$,
we obtain
\begin{eqnarray}
& D_{10} - D_{20} = \rho_{S} , \;\;\;
E_{1\pm} - E_{2\pm} = 0 , & \nonumber \\
& B_{10} - B_{20} = 0 , \;\;\; H_{1\pm} - H_{2\pm} = \mp i
J_{S\pm} . & \label{boundary}
\end{eqnarray}

To illustrate the use of the above formulas for static fields, we
choose the example of the electric field due to a ring having
radius $a$ with total charge $Q$ uniformly distributed and lying
in the $x$-$y$ plane, which is also worked out in
Ref.~\cite{Matute}. The surface charge density on $r=a$, localized
at $\theta = \pi / 2$, is
\begin{equation}
\rho_{S} = \frac{Q}{2 \pi a^2} \; \delta(\cos \, \theta) ,
\label{source1}
\end{equation}
which can be expanded using the series representation of the Dirac
delta function in terms of spherical harmonics
\begin{equation}
\delta(\cos \, \theta) = 2 \pi \sum_{l=0}^{\infty} \;
Y_{l0}(\frac{\pi}{2},0) \; Y_{l0}(\theta,0) .
\label{delta1}
\end{equation}
Taking into account the cylindrical symmetry of the system and the
requirement that the series solutions in Eqs.~(\ref{Laplace}) and
(\ref{v}) be finite at the origin, vanish at infinity and satisfy
the boundary conditions of Eq.~(\ref{boundary}) at $r=a$ for all
values of the angle $\theta$, namely, $E_{\pm}$ continuous at
$r=a$ and $E_{0}$ discontinuous at $r=a$, it is found that the
spin-weighted components of the electric field are given by
\begin{eqnarray}
E_{0} & = & \frac{Q}{\epsilon_{\circ} r^2} \sum_{l=0}^{\infty}
\frac{1}{2l+1} \; Y_{l0}(\frac{\pi}{2},0) \; Y_{l0}(\theta,0)
\nonumber \\ & & \times \left\{
\begin{array}{l}
        \displaystyle (l+1) \left( \frac{a}{r} \right)^{l}
        \; , \; r > a \\  \\
        \displaystyle  - l \left( \frac{r}{a} \right)^{l+1}
        \; , \; r < a
        \end{array}
\right.
\label{E0}
\end{eqnarray}
\begin{eqnarray}
E_{\pm} & = & \pm \frac{Q}{\epsilon_{\circ} r^2}
\sum_{l=1}^{\infty} \frac{\sqrt{l(l+1)}}{\sqrt{2}(2l+1)} \;
Y_{l0}(\frac{\pi}{2},0) \; _{\pm1}Y_{l0}(\theta,0) \nonumber \\
& & \nonumber \\ & & \times \left\{
\begin{array}{l}
        \displaystyle \left( \frac{a}{r} \right)^{l}
        \; , \; r > a \\  \\
        \displaystyle \left( \frac{r}{a} \right)^{l+1}
        \; , \; r < a
        \end{array}
\right.
\label{E+-}
\end{eqnarray}
Note that the discontinuity of the $l$th component of $E_{0}$ in
Eq.~(\ref{E0}) at $r=a$ is connected, according to
Eq.~(\ref{boundary}), with the corresponding component of the
surface charge density $\rho_{S}$ obtained from
Eqs.~(\ref{source1}) and (\ref{delta1}), exhibiting the unity of
the multipole expansions of fields and sources.

As an example of time-varying fields, we consider the problem of
the magnetic induction field from a current $I=I_{\circ} e^{-i
\omega t}$ in a circular loop of radius $a$ lying in the $x$-$y$
plane. The surface current density on $r=a$ is
\begin{equation}
{\bf J}_{S} = \frac{I_{\circ}}{a} \;
\delta(\cos \, \theta) \; e^{-i \omega t} \; \hat{\bf \varphi} ,
\label{source2}
\end{equation}
where for the delta function we now  use the expansion
\begin{equation}
\delta(\cos \, \theta) = 2 \pi \sum_{l=1}^{\infty} \;
_{\pm1}Y_{l0}(\frac{\pi}{2},0) \; _{\pm1}Y_{l0}(\theta,0) .
\label{delta2}
\end{equation}
The solution of the Helmholtz equation for the magnetic induction
field in Eqs. (\ref{Helmholtz}) and (\ref{w}), which is finite at
the origin, represents outgoing waves at infinity and satisfies
the boundary conditions of Eq.~(\ref{boundary}) at $r=a$ with
$J_{S\pm}={\bf e}_{\pm}\cdot{\bf J}_{S}$, becomes
\begin{eqnarray}
B_{0} & = & \pm i \frac{2\pi\mu_{\circ} I_{\circ} k a}{r} e^{-i
\omega t} \sum_{l=1}^{\infty} \sqrt{l(l+1)} \;
_{\pm1}Y_{l0}(\frac{\pi}{2},0) \nonumber \\ & & \times \;
Y_{l0}(\theta,0) \left\{
\begin{array}{l}
        \displaystyle j_{l}(ka) \; h_{l}^{(1)}(kr)
        \; , \; r > a \\  \\
        \displaystyle j_{l}(kr) \; h_{l}^{(1)}(ka)
        \; , \; r < a
        \end{array}
\right.
\label{B0}
\end{eqnarray}
\begin{eqnarray}
B_{\pm} & = & \displaystyle -i \frac{2\pi\mu_{\circ} I_{\circ}
k^{2} a}{\sqrt{2}} e^{-i \omega t} \sum_{l=1}^{\infty} \;
_{\pm1}Y_{l0}(\frac{\pi}{2},0) \; _{\pm1}Y_{l0}(\theta,0)
\nonumber \\ & & \times \left\{
\begin{array}{l}
        j_{l}(ka) \left[ h_{l-1}^{(1)}(kr) - \displaystyle
        \frac{l}{kr} h_{l}^{(1)}(kr) \right]
        \; , \; r > a \\  \\
        h_{l}^{(1)}(ka) \left[ j_{l-1}(kr) - \displaystyle
        \frac{l}{kr} j_{l}(kr) \right]
        \; , \; r < a
        \end{array}
\right.
\label{B+-}
\end{eqnarray}
The discontinuity of the $l$th component of $B_{\pm}$ in
Eq.~(\ref{B+-}) at $r=a$ is connected, according to
Eq.~(\ref{boundary}), with the $l$th component of the surface
current density $J_{S \pm}$ deduced from Eqs.~(\ref{source2}) and
(\ref{delta2}).

Finally, by using the expressions of the spin-weighted spherical
harmonics in Eqs.~(\ref{E0}), (\ref{E+-}), (\ref{B0}) and
(\ref{B+-}), it is seen that the results in Ref.~\cite{Matute} are
obtained.

We would like to thank G.F. Torres del Castillo for bringing to
notice Refs.~\cite{Torres1} and \cite{Torres2}. This work was
partially supported by Dicyt-Usach.

\end{document}